\documentclass[12pt]{article}
\usepackage{pic03}
\usepackage{hyperref}
\usepackage{url}
\usepackage{graphicx}

\begin{document}

\title{\bf CMS BARREL RESISTIVE PLATE CHAMBERS - TESTS AND RESULTS}

\author{ 
Borislav  Pavlov\\
 on behalf of the CMS RPC Barrel Collaboration\\
  Bari(Italy), Naples (Italy), Pavia (Italy), \\
  Peking University (China), Sofia University (Bulgaria), INRNE (Bulgaria)\\}

\maketitle

\baselineskip=14.5pt
\begin{abstract}
CMS experiment will use Resistive Plate Chambers (RPCs) as dedicated muon trigger detectors. The first 24 RPCs for the barrel muon system of the CMS experiment have been assembled and tested. A brief description of the cosmic muon telescope used for the tests is given. The preliminary results for the chamber performance are presented.
\end{abstract}

\baselineskip=17pt
\section{Introduction}
CMS (Compact Muon Solenoid) is a general purpose detector for the new coming LHC accelerator at CERN. Detailed description of CMS can be found elsewhere \cite{CMS_Proposal}.\\
RPCs \cite{RPC} are gaseous parallel-plate detectors with time resolution comparable to that of scintillators ($\sim$ 1 ns) and good spatial resolution ($\sim$ cm)\cite{CMS_Proposal}.\\
An RPC single gap consists of two parallel plates made of Bakelite and is filled with gas. The outer Bakelite surfaces are coated with graphite to form HV and ground electrodes. Two single gaps glued together form the double gap structure. The copper read-out strips are situated between the single gaps and are separated from the graphite coating by an insulating film. \\
A charged particle passing through an RPC initiates an electron avalanche due to ionisation in the applied electric field. The drift of electrons towards the anode induces a ``fast'' charge (signal) on the pick-up electrode. The developed more complex front-end electronics \cite{FEB} allows to operate the RPCs in avalanche mode \cite{aval1}.\\
To increase the signal (at a fixed voltage) or to reduce the voltage (at a fixed signal), double gap design is implemented by CMS (two gas gaps are mechanically assembled together in one parallel double gap structure with common read-out strips) \cite{aval1}, \cite{dg2}.\\
Prototype RPCs have already been build and successfully tested \cite{aval1}, \cite{dg2}. The results presented in this paper refers to double gaps design. 

\section{Test Set-Up}
The first 24 RPCs were tested using cosmic muons in Bari. The test set-up consists of the following systems:\\
{\bf Trigger System.} The RPCs are situated between two upper and two lower large trigger scintillators. The trigger signal is formed by taking the logical ``AND'' of the four large scintillators. Two additional small ``veto'' scintillators are used in order to suppress the contribution from electromagnetic avalanches. The chambers under tests have two double gaps (``forward'' and ``backward'') which are tested separately.\\
{\bf HV System.} Chamber's up and down single gaps are supplied separately.\\
{\bf DAQ System.} The front-end boards accepts signals from the chamber electrodes and after amplification and discrimination forms a LVDS signal for every strip. RPC signals are read-out at a fixed frequency of 40 MHz and are stored in a temporary memory. When the trigger arrives the information is passed to the PC for further processing.\\
{\bf Gas System.} The gas system allow as to use up to four component gas mixture.

\begin{figure}[htb]
\begin{center}
\scalebox {0.18}[0.35]{\includegraphics{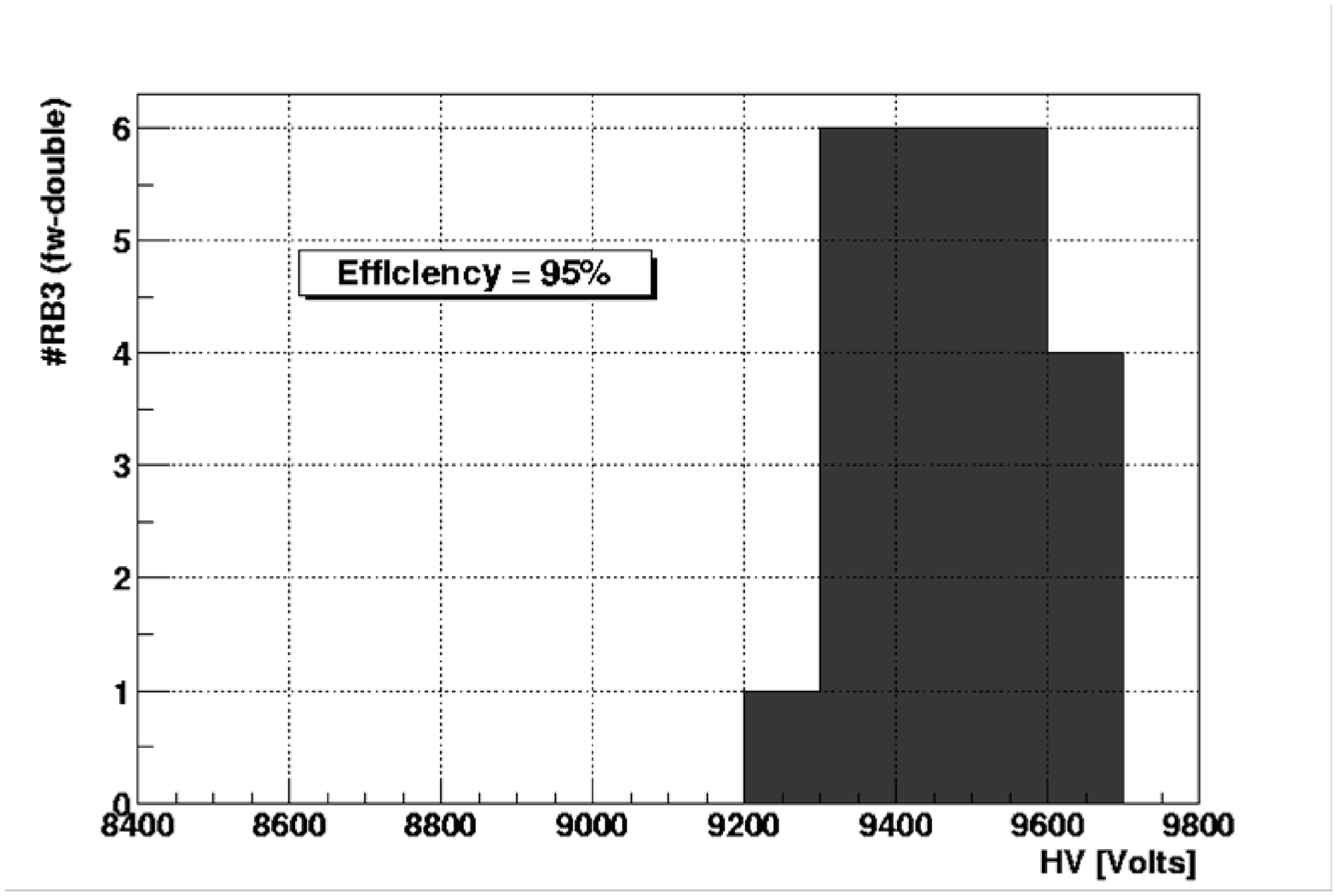}}
\scalebox {0.23}[0.35]{\includegraphics{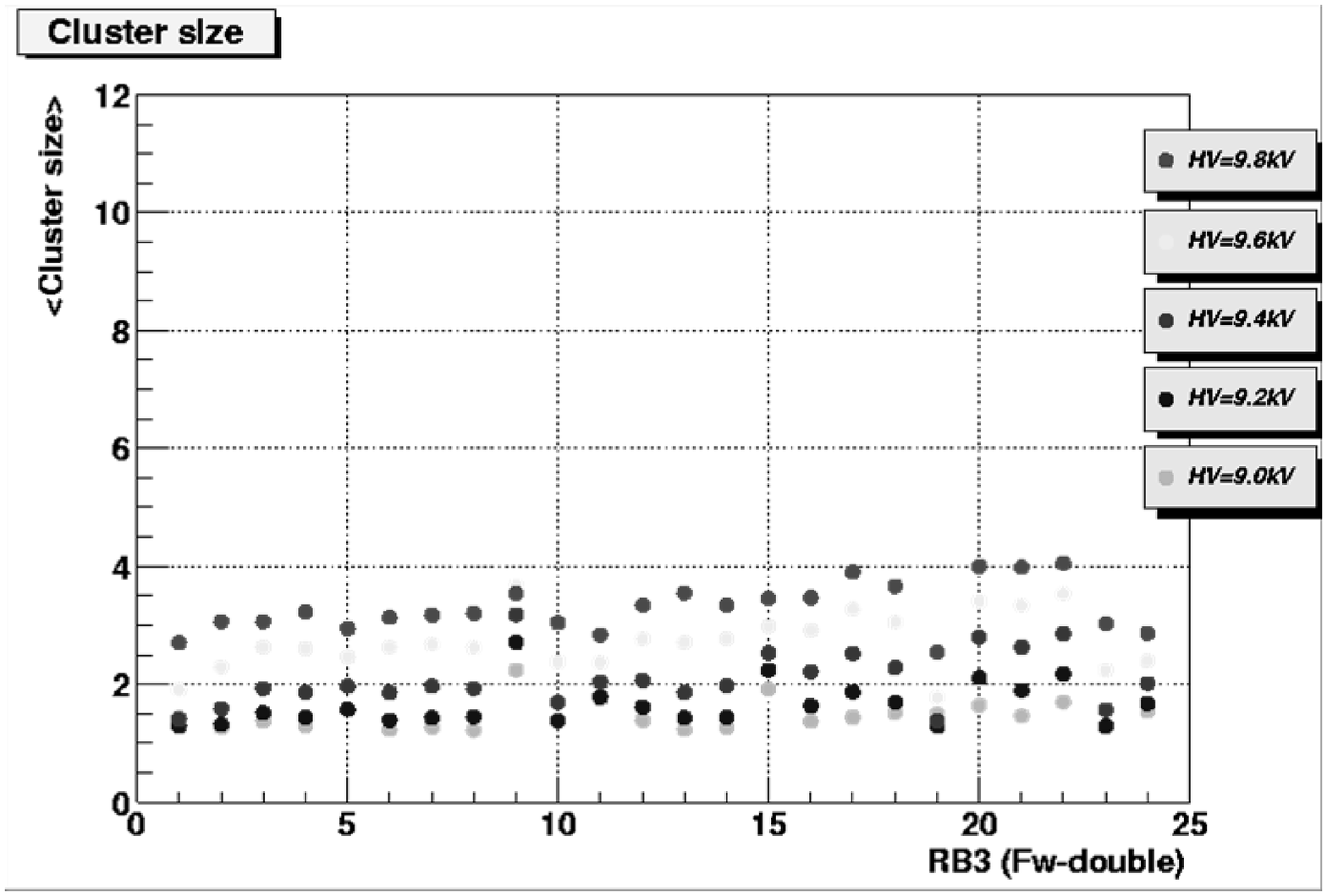}}
\scalebox {0.23}[0.35]{\includegraphics{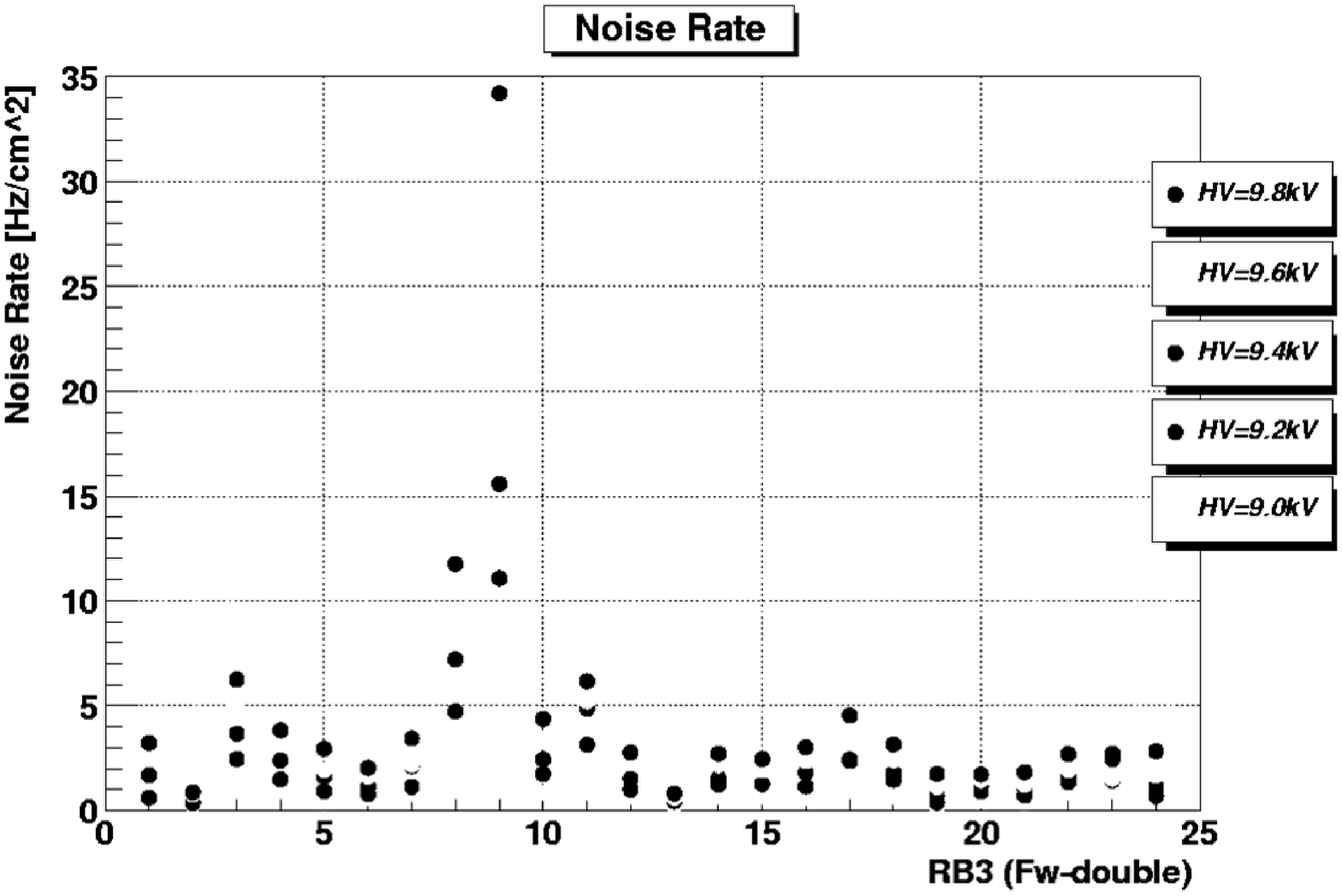}}
\caption{\it Distribution of the 95\% efficiency voltage (left). Cluster size (middle) and noise rate (right) versus the chamber number for different voltages. \label{effclsnoise} }
\end{center}
\end{figure}

\section{RPC tests and results}
The efficiency, cluster size, noise rate and electric current were measured for different voltages in the range 9.0 - 9.8 kV. The chamber front-end electronics threshold was fixed during the tests to accept the charge above 125 fC. Temperature and atmospheric pressure corrections have been made. Freon-based gas mixture has been used (96.2\% $C_2$$H_2$$F_4$, 3.5\% i-$C_4$$H_{10}$, 0.3\% i-S$F_{6}$).\\
{\bf Efficiency.} The efficiency is defined as a ratio between the RPC responses (at least one fired strip) and the number of recorded triggers, with correction for spurious hits. All chambers have achieved desired efficiency of 95\% and above (fig. \ref{effclsnoise}). Long and stable efficiency plateaus are observed.\\
{\bf Cluster size.} Cluster size is defined as the number of simultaneously fired adjacent strips. The average cluster size for the voltages in the range 9.4-9.6 kV is between 1.5 and 3 (fig. \ref{effclsnoise}). This is higher than expected, but is still acceptable. The cluster size can be reduced by changing the front-end electronics thresholds to optimal value.\\
{\bf Noise rate.} Noise rate is defined as the total chamber noise normalised to its active surface. A noise rate of up to 10 Hz/$cm^{2}$ is considered acceptable within the detector specifications. However, the test results show an even improved noise rate of less than 5 Hz/$cm^{2}$, (fig. \ref{effclsnoise}). Only one chamber has unacceptable noise rate (CH 9) and will be reassembled.
\section{Conclusions}
The first 24 RPC chambers for muon station RB3 were produced and tested. 
All of them reach the desired high efficiency (above 95\%). 
The cluster size and noise rate at high voltages 9.4-9.6 kV fully satisfy CMS requirements (except CH 9).


\begin{thebibliography}{99}
\small
\bibitem{CMS_Proposal} CMS collaboration, Technical Proposal, CERN/LHCC 94-38, (1994)
\bibitem{RPC} R. Santonico {\it et al}, NIM {\bf 187}, 377 (1981)
\bibitem{FEB} M. Abbrescia {\it et al}, New developments on front-end electronics for the CMS Resistive Plate Chambers NIM A {\bf 456}, 143-149 (2000)
\bibitem{aval1} M. Abbrescia {\it et al}, Local and global performance of double gap RPCs operated in avalanche mode, NIM A {\bf 434}, 244-253 (1999) 
\bibitem{dg2} M. Abbrescia {\it et al}, Recent result and developments on double-gap RPC for CMS, NIM A {\bf 409}, 43-45 (1998)
\end{thebibliography}
\end{document}